\RequirePackage{lineno}
\documentclass[aps,prl,superscriptaddress,showpacs,twocolumn]{revtex4-1}
\usepackage{graphicx}
\usepackage{floatflt}
\usepackage{float}
\usepackage{tabularx}
\usepackage{comment}

\def\Ag{A^{np}_\gamma}

\def\18F{^{18}F}
\def\DI{\Delta I}

\def\P-odd{P-{\rm{odd}}}

\def\gtorder{\mathrel{\raise.3ex\hbox{$>$}\mkern-14mu
 \lower0.6ex\hbox{$\sim$}}}
\def\ltorder{\mathrel{\raise.3ex\hbox{$<$}\mkern-14mu
 \lower0.6ex\hbox{$\sim$}}}

\begin{document}
\title{First Observation of $P$-odd $\gamma$ Asymmetry in Polarized Neutron Capture on Hydrogen}


\author{D.~Blyth}
\affiliation{Arizona State University, Tempe, AZ 85287}
\affiliation{High Energy Physics Division, Argonne National Laboratory, Argonne, IL, 60439, USA}
\author{J.~Fry}
\affiliation{University of Virginia, Charlottesville, VA 22904, USA}
\affiliation{Indiana University, Bloomington, IN 47405, USA}
\author{N.~Fomin}
\affiliation{University of Tennessee, Knoxville, TN 37996, USA}
\affiliation{Los Alamos National Laboratory, Los Alamos, NM 87545, USA}
\author{R.~Alarcon}
\affiliation{Arizona State University, Tempe, AZ 85287}
\author{L.~Alonzi}
\affiliation{University of Virginia, Charlottesville, VA 22904, USA}
\author{E.~Askanazi}
\affiliation{University of Virginia, Charlottesville, VA 22904, USA}
\author{S.~Bae{\ss}ler}
\affiliation{University of Virginia, Charlottesville, VA 22904, USA}
\affiliation{Oak Ridge National Laboratory, Oak Ridge, TN 37831, USA}
\author{S.~Balascuta}
\affiliation{Horia Hulubei National Institute for Physics and Nuclear Engineering, Magurele 077125, Romania}
\affiliation{Arizona State University, Tempe, AZ 85287}
\author{L.~Barr\'{o}n-Palos}
\affiliation{Instituto de F\'{i}sica, Universidad Nacional Aut\'{o}noma de M\'{e}xico, Apartado Postal 20-364, 01000, M\'{e}xico}
\author{A.~Barzilov}
\affiliation{University of Nevada, Las Vegas, NV 89154, USA}
\author{J.D.~Bowman}
\affiliation{Oak Ridge National Laboratory, Oak Ridge, TN 37831, USA}
\author{N.~Birge}
\affiliation{University of Tennessee, Knoxville, TN 37996, USA}
\author{J.R.~Calarco}
\affiliation{University of New Hampshire, Durham, NH 03824, USA}
\author{T.E.~Chupp}
\affiliation{University of Michigan, Ann Arbor, MI 48109, USA}
\author{V.~Cianciolo}
\affiliation{Oak Ridge National Laboratory, Oak Ridge, TN 37831, USA}
\author{C.E.~Coppola}
\affiliation{University of Tennessee, Knoxville, TN 37996, USA}
\author{C.~B.~Crawford}
\affiliation{University of Kentucky, Lexington, KY 40506, USA}
\author{K.~Craycraft}
\affiliation{University of Tennessee, Knoxville, TN 37996, USA}
\affiliation{University of Kentucky, Lexington, KY 40506, USA}
\author{D.~Evans}
\affiliation{University of Virginia, Charlottesville, VA 22904, USA}
\affiliation{Indiana University, Bloomington, IN 47405, USA}
\author{C.~Fieseler}
\affiliation{University of Kentucky, Lexington, KY 40506, USA}
\author{E.~Frle\v{z}}
\affiliation{University of Virginia, Charlottesville, VA 22904, USA}
\author{I.~Garishvili}
\affiliation{Oak Ridge National Laboratory, Oak Ridge, TN 37831, USA}
\affiliation{University of Tennessee, Knoxville, TN 37996, USA} 
\author{M.T.W.~Gericke}
\affiliation{University of Manitoba, Winnipeg, MB, Canada R3T 2N2}
\author{R.C.~Gillis}
\affiliation{Oak Ridge National Laboratory, Oak Ridge, TN 37831, USA}
\affiliation{Indiana University, Bloomington, IN 47405, USA}
\author{K.B.~Grammer}
\affiliation{Oak Ridge National Laboratory, Oak Ridge, TN 37831, USA}
\affiliation{University of Tennessee, Knoxville, TN 37996, USA}
\author{G.L.~Greene}
\affiliation{University of Tennessee, Knoxville, TN 37996, USA}
\affiliation{Oak Ridge National Laboratory, Oak Ridge, TN 37831, USA}
\author{J.~Hall}
\affiliation{University of Virginia, Charlottesville, VA 22904, USA}
\author{J.~Hamblen}
\affiliation{University of Tennessee, Chattanooga, TN 37403 USA}
\author{C.~Hayes}
\affiliation{Physics Department, North Carolina State University, Raleigh, NC 27695, USA}
\affiliation{University of Tennessee, Knoxville, TN 37996, USA}
\author{E.B.~Iverson}
\affiliation{Oak Ridge National Laboratory, Oak Ridge, TN 37831, USA}
\author{M.L.~Kabir}
\affiliation{Mississippi State University, Mississippi State, MS 39759, USA}
\affiliation{University of Kentucky, Lexington, KY 40506, USA}
\author{S.~Kucuker}
\affiliation{Northwestern University Feinberg School of Medicine, Chicago, IL 60611, USA}
\affiliation{University of Tennessee, Knoxville, TN 37996, USA}
\author{B.~Lauss}
\affiliation{Paul Scherrer Institut, CH-5232 Villigen, Switzerland}
\author{R.~Mahurin}
\affiliation{Middle Tennessee State University, Murfreesboro, TN 37132, USA}
\author{M.~McCrea}
\affiliation{University of Kentucky, Lexington, KY 40506, USA}
\affiliation{University of Manitoba, Winnipeg, MB, Canada R3T 2N2}
\author{M. Maldonado-Vel\'{a}zquez}
\affiliation{Instituto de F\'{i}sica, Universidad Nacional Aut\'{o}noma de M\'{e}xico, Apartado Postal 20-364, 01000, M\'{e}xico}
\author{Y.~Masuda}
\affiliation{High Energy Accelerator Research Organization (KEK), Tukuba-shi, 305-0801, Japan}
\author{J.~Mei}
\affiliation{Indiana University, Bloomington, IN 47405, USA}
\author{R.~Milburn}
\affiliation{University of Kentucky, Lexington, KY 40506, USA}
\author{P.E.~Mueller}
\affiliation{Oak Ridge National Laboratory, Oak Ridge, TN 37831, USA}
\author{M.~Musgrave}
\affiliation{Massachusetts Institute of Technology, Cambridge, MA 02139, USA}
\affiliation{University of Tennessee, Knoxville, TN 37996, USA}
\author{H.~Nann}
\affiliation{Indiana University, Bloomington, IN 47405, USA}
\author{I.~Novikov}
\affiliation{Western Kentucky University, Bowling Green, KY 42101, USA}
\author{D.~Parsons}
\affiliation{University of Tennessee, Chattanooga, TN 37403 USA}
\author{S.I.~Penttil\"a}
\affiliation{Oak Ridge National Laboratory, Oak Ridge, TN 37831, USA}
\author{D.~Po\v{c}ani\'{c}}
\affiliation{University of Virginia, Charlottesville, VA 22904, USA}
\author{A.~Ramirez-Morales}
\affiliation{Instituto de F\'{i}sica, Universidad Nacional Aut\'{o}noma de M\'{e}xico, Apartado Postal 20-364, 01000, M\'{e}xico}
\author{M.~Root}
\affiliation{University of Virginia, Charlottesville, VA 22904, USA}
\author{A.~Salas-Bacci}
\affiliation{University of Virginia, Charlottesville, VA 22904, USA}
\author{S.~Santra} 
\affiliation{Bhabha Atomic Research Centre, Trombay, Mumbai 400085, India}
\author{S.~Schr\"{o}der}
\affiliation{University of Virginia, Charlottesville, VA 22904, USA}
\affiliation{Saarland University, Institute of Experimental Ophthalmology, Kirrberger Str. 100, Bldg. 22, 66424 Homburg/Saar, Germany}
\author{E.~Scott}
\affiliation{University of Tennessee, Knoxville, TN 37996, USA}
\author{P.-N. Seo}
\affiliation{University of Virginia, Charlottesville, VA 22904, USA}
\affiliation{Triangle Universities Nuclear Lab, Durham, NC 27708, USA}
\author{E.I.~Sharapov}
\affiliation{Joint Institute for Nuclear Research, Dubna 141980, Russia}
\author{F.~Simmons}
\affiliation{University of Kentucky, Lexington, KY 40506, USA}
\author{W.M.~Snow}
\affiliation{Indiana University, Bloomington, IN 47405, USA}
\author{A.~Sprow}
\affiliation{University of Kentucky, Lexington, KY 40506, USA}
\author{J.~Stewart}
\affiliation{University of Tennessee, Chattanooga, TN 37403 USA}
\author{E.~Tang}
\affiliation{University of Kentucky, Lexington, KY 40506, USA}
\affiliation{Los Alamos National Laboratory, Los Alamos, NM 87545, USA}
\author{Z.~Tang}
\affiliation{Indiana University, Bloomington, IN 47405, USA}
\affiliation{Los Alamos National Laboratory, Los Alamos, NM 87545, USA}
\author{X.~Tong}
\affiliation{Oak Ridge National Laboratory, Oak Ridge, TN 37831, USA}
\author{D.J.~Turkoglu}
\affiliation{National Institute of Standards and Technology, Gaithersburg, MD 20899, USA}
\author{R.~Whitehead}
\affiliation{University of Tennessee, Knoxville, TN 37996, USA}
\author{W.S.~Wilburn}
\affiliation{Los Alamos National Laboratory, Los Alamos, NM 87545, USA}

\collaboration{The NPDGamma Collaboration}

\date{\today}

\begin{abstract}

\textit{We report the first observation of the parity-violating gamma-ray asymmetry $\Ag$ in neutron-proton capture using polarized cold neutrons incident on a liquid parahydrogen target at the Spallation Neutron Source at Oak Ridge National Laboratory. $\Ag$ isolates the $\DI=1$,  $^{3}S_{1}\rightarrow^{3}P_{1}$ component of the weak nucleon-nucleon interaction, which is dominated by pion exchange and can be directly related 
to a single coupling constant in either the DDH meson exchange model or pionless effective field theory. We measured $\Ag = (-3.0 \pm 1.4 (stat.) \pm 0.2 (sys.))\times 10^{-8}$,
which implies a DDH weak $\pi NN$ coupling of $h^1_\pi=(2.6 \pm 1.2 (stat.) \pm 0.2 (sys.))\times10^{-7}$
and a pionless EFT constant of $C^{^{3}S_{1}\rightarrow ^{3}P_{1}}/C_{0}=(-7.4 \pm 3.5 (stat.) \pm 0.5(sys.)) \times 10^{-11}$  MeV$^{-1}$. 
We describe the experiment, data analysis, systematic uncertainties, and implications of the result.}


\end{abstract}


\maketitle


\paragraph{Introduction.}

In this Letter we present the first observation of the parity-violating (PV) asymmetry $\Ag$ of gammas emitted from the capture of polarized neutrons on protons. Analysis of the asymmetry leads to the first determination of an isolated term in the weak nucleon-nucleon (NN) potential. This represents a major step toward a complete experimental determination of the spin-isospin structure of the hadronic weak interaction (HWI). 

The electroweak component of the standard model (SM) describes the weak couplings of $W^{\pm}$ and $Z$ gauge bosons to quarks and, in principle, the HWI. The HWI causes parity-violating admixtures in nuclear wave functions and produces small but observable PV spin-momentum correlations and photon circular polarizations. However, nonperturbative QCD dynamics make a direct calculation of PV nuclear observables out of reach.

Desplanques, Donoghue, and Holstein (DDH)~\cite{Desplanques1980} introduced a meson exchange model to describe the HWI. This model is parametrized by six parity-odd time-reversal-even rotational invariants that can be constructed from the spin, isospin, momenta, and coordinates of the interacting nucleons. Each term has a Yukawa dependence in the separation of the nucleons with range determined by the mass of the exchanged meson ($\pi$, $\rho$, or $\omega$). The six adjustable coupling constants are labeled by the meson exchanged and the change of the total isospin $\DI$: $h_{\pi}^1$, $h_{\rho}^{0,1,2}$, and $h_{\omega}^{0,1}$. DDH also give reasonable ranges for these coupling constants. Observables are calculated as matrix elements of the PV potential terms between nuclear states and the coupling constants are to be determined from experiment.

The two-body $n$-$p$ system is exactly calculable once the strong NN interaction is specified and there is no nuclear structure uncertainty in the interpretation of $\Ag$. $\Ag$ depends on only $\DI=1$ coupling constants.
Similarly, the value of the circular polarization, $P_{\gamma}$, of the 1.081 MeV $\gamma$ emitted by unpolarized $^{18}$F nuclei~\cite{four18F} depends only on the $\DI=1$ terms in the HWI. However, the contributions from heavy meson terms are much larger in $P_{\gamma}$ than in $\Ag$ allowing a determination of $h_{\pi}^1$ and a linear combination of $\DI=1$ heavy meson couplings in a combined analysis. 

New theoretical approaches to weak NN interactions based on effective field theory (EFT) and the $1/N_{c}$ expansion of QCD, where $N_{c}$ is the number of colors, predict relative sizes of PV couplings. In pionless EFT, the HWI is described by five $S$-$P$ transition amplitudes first introduced by Danilov~\cite{Dan65} and elaborated in subsequent work~\cite{Dan72, Zhu2005, Phillips2009, Schindler2013}. In the pionless EFT approach~\cite{Schindler2013}, $\Ag$ is proportional to the $\DI=1$ low energy constant $C^{^{3}S_{1}\rightarrow ^{3}P_{1}}/C_{0}$. Recently the $1/N_{c}$ expansion of QCD~\cite{tHooft74, Witten79, Jenkins1998, Cohen2012, DeGrand2016} has been applied to the HWI. Phillips {\it et al.}~\cite{Phillips2015, Samart2016} constructed the $1/N_{c}$ expansion of the DDH couplings and Schindler {\it et al.}~\cite{Schindler2016} have developed the $1/N_{c}$ expansion in pionless EFT, valid for two-body systems at low energy, and the phenomenology was analyzed by Gardner {\it et al.}~\cite{Gardner17}. In addition to $1/N_{c}$ dependence, all $\DI=1$ terms in both DDH and EFT theories are suppressed by a factor $\sin^{2}(\theta_{W})=0.223$. Since charged currents are suppressed in $\Delta I = 1$ NN processes by ${V_{us}^{2}/V_{ud}^{2}}=0.053$, the weak NN interaction is one of the few systems sensitive to quark-quark neutral current effects~\cite{Adelberger1985,Haxton13}. Within each of the different theoretical approaches described above, predictions for the relative size of weak NN amplitudes in different meson and isospin channels vary by an order of magnitude. Their relative sizes may reveal new aspects of strong QCD, and their calculation within the SM has consequently been the subject of extensive theoretical work~\cite{Khatsimovsky1985,Dubovik1986,Kaiser1988,Kaiser1990,FCDH,Henley1996,Henley1998,Meissner1999,Zhu2001, Savage2001, QCDSum,  Liu2007,Hyun2007,Hyun2008,Desplanques2008, Gazit2008, Schindler2010,Lee2012,wasem,Vries2013,Vivani2014, Vries2014, deVries2015pza, Feng2018,Hyun2005}. Finally, lattice gauge theory calculations present an exciting intellectual opportunity for understanding nonperturbative aspects of QCD. Wasem~\cite{wasem} has published a pioneering lattice QCD calculation of the contribution of connected diagrams to $h_{\pi}^1$.

\begin{figure}[htpt] 
\includegraphics[width=0.45\textwidth]{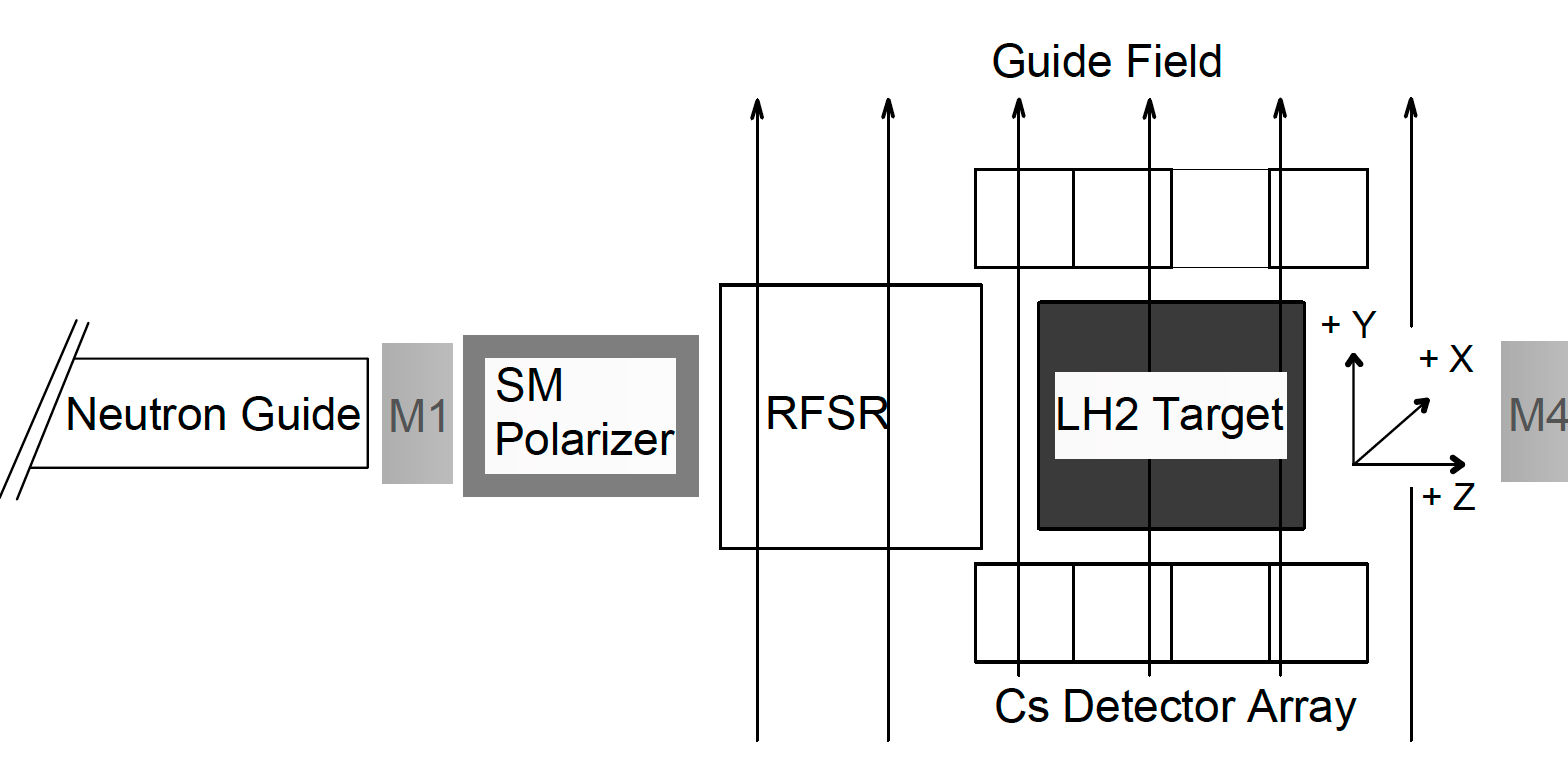}
\caption{A schematic vertical cut view of the NPDGamma experiment on the FnPB, for details see text.}
\label{npdg_diag1}
\end{figure}

\paragraph{Experiment.}
We measured $\Ag$ on the fundamental neutron physics beamline (FnPB) at the spallation neutron source (SNS) using the same apparatus as the first phase of the experiment ~\cite{gericke} with some improvements.
At the SNS proton pulses delivered at 60 Hz to a mercury target produce spallation neutrons which are cooled by a liquid hydrogen moderator. The neutrons travel 15~m down a supermirror (SM) neutron guide~\cite{nadia} to the NPDGamma experiment. Two choppers select neutron wavelengths between $3.1$-$6.6$ \AA~ from each 60 Hz time-of-flight (TOF) pulse and reject neutrons outside this range to prevent lower energy neutrons mixing into the next pulse. The neutron beam intensity was sampled by two $^{3}$He ionization chambers, one upstream (M1) and one downstream (M4) from the hydrogen target~\cite{kyle,gericke}, see Fig.~\ref{npdg_diag1}. M1 absorbed approximately $1\%$ of the beam and determined the number of neutrons in each pulse with a statistical uncertainty of $10^{-4}$.
%
%


After M1, neutrons passed through a SM polarizer and emerged with an average polarization of 94\%~\cite{matt}. The neutron spin was transported to the target by a uniform magnetic field $\vec{B_0} = 9.5$~G aligned within 3 mrad to the +$\hat{y}$ axis. To eliminate Stern-Gerlach beam steering, the gradient was limited to $\partial B_y/\partial y \leq 2$~mG/cm within the volume between the RF Spin Rotator (RFSR) and the target volume~\cite{septimiu, pil}. The neutron flux at the LH$_2$ target position was $7.7 \times 10^{9}$ n/s at 1 MW~\cite{nadia,Fry}.


$\Ag$ was determined from interactions of the polarized neutron beam on a 16~l liquid hydrogen (LH$_2$) target in the parahydrogen ({\it p}-H$_2$) molecular state~\cite{kyle,santra}. Scattering from the $S=0$ {\it p}-H$_2$ molecular ground state preserves neutron polarization for incident neutron energies which fall below the $14.7$ meV threshold for spin-flip scattering into the $S=1$ orthohydrogen ({\it o}-H$_2$) molecular ground state. 
The {\it o}-H$_2$ fraction $f_{\rm {o-H_2}}$, which can flip the neutron spin upon scattering, was minimized by continuously circulating the liquid through a catalytic converter operated at 15.4~K~\cite{kyle}. 
Because of the long neutron mean free path in {\it p}-H$_2$, only about 43\% of the incident 
neutrons were captured by {\it p}-H$_2$. The rest 
were scattered by the LH$_2$ and absorbed by the target vessel made from an aluminum alloy or 
by a $^{6}$Li-loaded neutron absorber wrapped on the outside surface of the vessel. 
$f_{\rm {o-H_2}}$ was monitored periodically with neutron transmission measurements using M1 and M2~\cite{kyle}.
We measured the neutron-{\it p}-H$_2$ scattering cross sections and used that to determined an upper limit of $f_{\rm {o-H_2}}<0.0015$~\cite{kyle}. With this limit, we estimated the neutron depolarization to be $0.032 \pm 0.016$ using MCNPX~\cite{Pelowitz2011} and the cross sections in Ref. ~\cite{Young}.  

$\gamma$ rays were detected with an array of 48 cubical CsI(Tl) detectors (sides 15.2 cm) arranged symmetrically in four rings of 12 covering $\approx$ 3$\pi$ sr \cite{gericke,gericke2005}. 
The detector array was aligned within 3~mrad to the local magnetic field direction 
to suppress any mixing of the PV (up-down) asymmetry with the parity-conserving (left-right) asymmetry~\cite{Csoto1997}.
The detectors were operated in current mode due to high instantaneous detector rates of $\sim 10^8$ Hz. Scintillation light was converted to a voltage signal using magnetic field insensitive vacuum photodiodes and low-noise amplifiers~\cite{gericke}. The spectral density of the amplifier noise was measured to be much smaller than the shot noise density from $\gamma$ counting statistics~\cite{michael2, wilburn}. The ability of the apparatus to detect a PV asymmetry was tested by measuring the large ($\sim 3 \times 10^{-5}$) PV $\gamma$ asymmetry from polarized slow neutron capture on $^{35}$Cl~\cite{Vesna82, Avenier1985, Mitchell2004}. We observed asymmetries consistent with previous work~\cite{Fomin2018}. 

The prompt signal from the LH$_2$ target consisted of $\sim$80\% $\gamma$'s from capture on hydrogen and $\sim$20\% $\gamma$'s from capture on aluminum. Neutrons that capture on $^{28}$Al produce a prompt PV $\gamma$ cascade, followed by a $\beta$-delayed $\gamma$ ($\tau$ = 194 s). The $\beta$-delayed signal manifests as a constant pedestal. The prompt PV $\gamma$ asymmetry in aluminum must be measured separately. The aluminum prompt $\gamma$ asymmetry was first measured using the same apparatus, replacing the LH$_2$ target with an aluminum target. The apparatus was then removed to allow for installation of the next experiment (n-$^3$He). During data analysis, the importance of constructing the aluminum target from the same material used to fabricate the LH$_2$ target vessel became clear. So, the apparatus was reinstalled to remeasure the aluminum asymmetry. The different aluminum components of the apparatus such as the RFSR windows, cryostat vacuum windows, target vessel entrance and exit windows, and vessel side walls could have different prompt $\gamma$ asymmetries due to different impurities. To account for this, we built 4 targets from the 4 different components of the apparatus and one target from the window material of the new RFSR. We also built one composite target that incorporated material from each component with mass proportional to their relative yields to the prompt signal, as determined by Monte Carlo calculation ~\cite{Blyth}. For these measurements, we used the improved DAQ and the high-efficiency RFSR from the $n$-$^{3}$He experiment. 

\paragraph{Data, analysis, and results.}

For each neutron pulse, the current-mode signals from each detector were digitized to give 40 time bins of differential photon yield. These differential yields were summed over a fiducial time interval for which both choppers were open and the neutron polarization was well defined for each spin direction $\uparrow\downarrow$. The neutron polarization was reversed with a 16-step spin sequence (SS) $\uparrow\downarrow\downarrow\uparrow\downarrow\uparrow\uparrow\downarrow$
$\downarrow\uparrow\uparrow\downarrow\uparrow\downarrow\downarrow\uparrow$. A total of 5.9$\times10^7$ SS were accumulated during the LH$_2$ running. This pattern rejects known 30 Hz beam intensity fluctuations and suppresses drifts up to 3rd order.

\begin{figure}[htpt]
\includegraphics[width=0.45\textwidth]{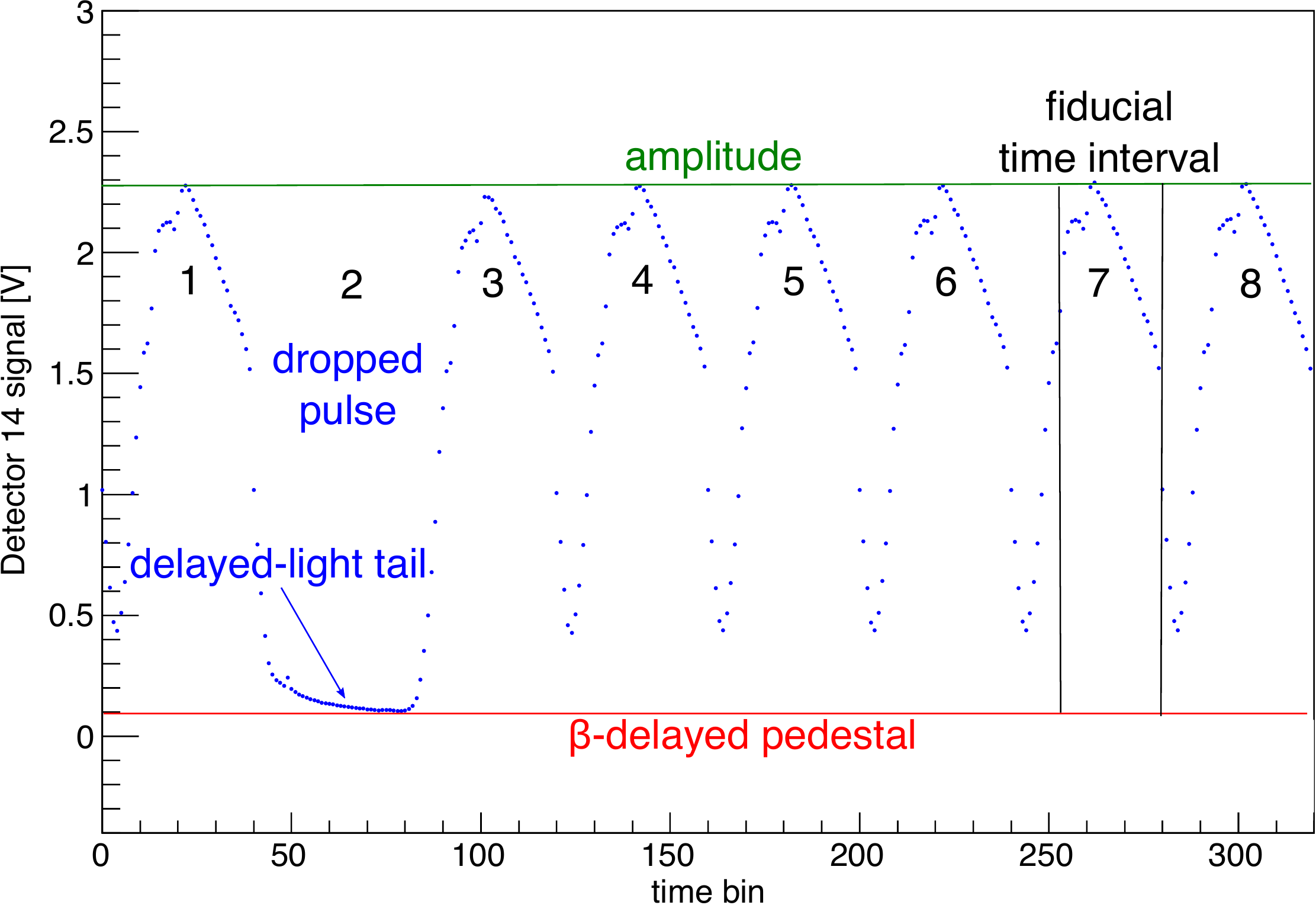}
\caption{Plot of a typical detector voltage signal as a function of time bin for eight 60 Hz neutron pulses. The proton pulse was not delivered to the spallation target in the $2^{nd}$ pulse resulting in a dropped pulse. The peak yield in the 3rd pulse is 1\% low because the phosphorescence tail from the second pulse is missing. The rising (falling) edges of the pulses correspond to the choppers opening (closing). The pedestal from the $\beta$-delayed $\gamma$s of $^{28}$Al is shown. Finally, the fiducial time interval (27 time bins wide) is shown in pulse seven (time bins 253 to 279).}
\label{droppedpulsePRL}
\end{figure}

The contributions to the detector yields must be understood to determine the PV asymmetries. The $\beta$-delayed $\gamma$s and small electronic offsets combine to form a pedestal that is nearly time independent on the scale of a SS. Each CsI(Tl) detector also has a delayed-light, multicomponent phosphorescence tail \cite{Phosph} with a typical decay time of $6.7 \pm 1.6$ ms contributing 1\% of the yield in the subsequent pulse (see Fig. ~\ref{droppedpulsePRL}). The tails are assumed to have the same PV and intensity variations as the prompt yields. The asymmetry for detector $d$ is defined in terms of prompt photon yields, $Y_d$, as  $A_{d}={{(Y_{d}^{\uparrow}-Y_{d}^{\downarrow})} \over {(Y_{d}^{\uparrow}+Y_{d}^{\downarrow})}}$, but is not measured directly. The measured detector yields contain nonprompt contributions (and delayed light tails) as defined above. These contributions can be determined from ``dropped pulses',' in which protons were not sent to the spallation target and the prompt photons are not present in the signal, but nonprompt contributions are (see Fig.~\ref{droppedpulsePRL}).  Three different analyses used information from dropped pulses to properly normalize the asymmetries.

All data for which the apparatus was operating normally were included in the analysis. Roughly 20\% of SS were eliminated because of unstable beam power, improper chopper phasing (which impacts the fiducial time window) or RFSR errors. The measured neutron intensity in the polarization-insensitive monitor M1 was used to apply the beam power cuts, which accounted for nearly all of the eliminated data. Figure 3 shows the effect of these cuts on the asymmetry  of a typical detector. After cuts were applied, the asymmetry distributions were indistinguishable from Gaussian~\cite{Fry2017}. The extracted asymmetries determined using three different analyses agreed to within a small fraction of the statistical uncertainties.

The aluminum asymmetry measurements were taken with a different DAQ and RFSR using a simple 30 Hz neutron spin state reversal pattern $\uparrow\downarrow\uparrow\downarrow \cdots$, with a total of 1.5$\times10^7$ SS accumulated. This simple reversal pattern introduced a sensitivity to a 30~Hz neutron intensity modulation of 10$^{-4}$.  Proper normalization of raw detector asymmetries was applied to remove detector dependence from such 30 Hz signals. 
The information needed to normalize the detector responses was determined from the detector yields in the neighborhood of the dropped pulses~\cite{Blyth,BlythAlPRC}. 
Detector-pair asymmetries were formed from the difference of azimuthally opposing detector asymmetries to extract the physics result. In order to verify that the normalization sufficiently suppressed the 30~Hz modulation, a regression analysis was performed between the beam intensity modulation extracted from $M$1 signals and the pair asymmetries. The slope of this regression was consistent with zero.

The differential cross section for the direction of the capture $\gamma$s with respect to the spin direction is ${d\sigma \over d\Omega} \sim 1+A_{\gamma} {\vec{k_{\gamma}} \cdot \vec{s_{n}}}$, neglecting parity-conserving contributions.
Correcting for the finite geometry of the beam, target, and detectors requires a Monte Carlo calculation of the energy-weighted values of the average scalar product $k_\gamma \cdot s_n$ for each detector, denoted ``geometric factors.'' The geometric factors are calculated for all $\gamma$ rays from simulated neutron capture in the target, target vessel, and its surrounding shielding which deposit energy in a detector element. Compton scattering causes a single $\gamma$ to deposit energy in more than one detector leading to correlations between energy depositions in different detectors. These correlations lead to non-diagonal uncertainty covariance matrices. The geometric factors were calculated using GEANT4 and MCNPX simulations~\cite{grammer2018gf, Blyth} and the covariances were determined from data.

The relationship between the pair asymmetries $A_p$ and the physics asymmetries $A_{\gamma}$ becomes \hfill \break
$A_{p}=\displaystyle\sum_{i} P_{\text{tot}}^if_{p}^iG_{p}^iA_{\gamma}^i$, where $P_{\text{tot}}^i$, $f_{p}^i$, $G_{p}^i$ and $A_{\gamma}^i$ are the net polarization factor (beam polarization, target depolarization, and RFSF efficiency), the fractional contribution to the detector yield, the geometric factor, and the $\gamma$ asymmetry of the $i$th target component (e.g., hydrogen, aluminum window, etc.) respectively, for detector pair $p$. 

\begin{figure}[t!]
\includegraphics[width=0.45\textwidth]{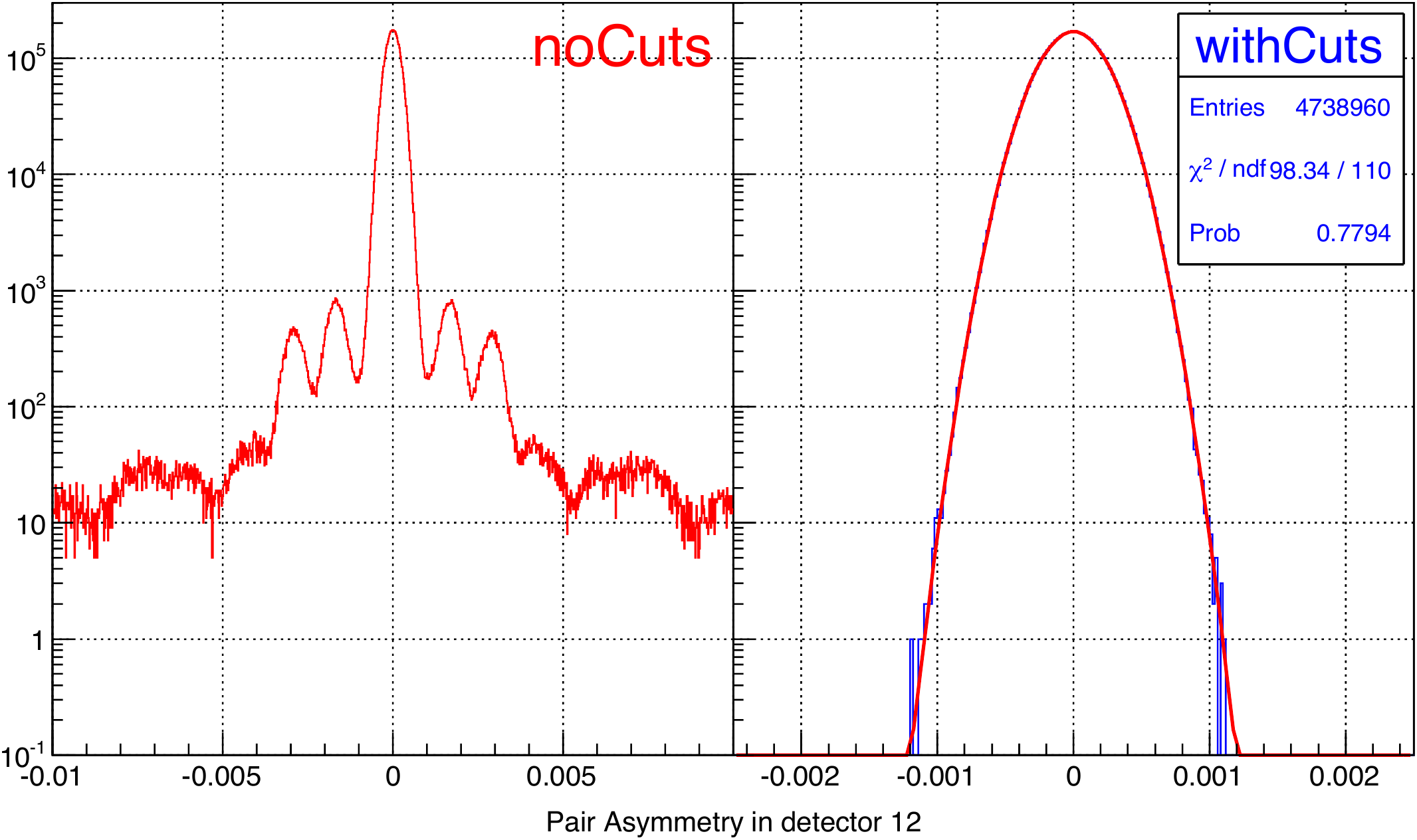}
\caption{Histogram of hydrogen asymmetries ($\sim$1/30 of all the data) for a typical detector before (left) and after (right) the cuts described in the text have been applied. Note the different $x$-axis scale on the right panel. The distinct side lobes in the uncut data correspond to SS in which one or more dropped pulses occurred.}
\label{pairs}
\end{figure}

The hydrogen and aluminum asymmetries were simultaneously extracted from a $\chi^2$ minimization scheme using data sets from hydrogen and aluminum targets as well as the corresponding sets of $P_{tot}^i$, $f_p^i$, and $G_p^i$. Three different analyses were consistent in their results. The integrated $\chi^2$ probability for each analysis was 0.73, 0.64, and 0.43. The extracted hydrogen asymmetry is $\Ag=[-3.0\pm 1.4(\text{stat})] \times 10^{-8}$ and the extracted aluminum PV asymmetry is $[-12 \pm 3(\text{stat})] \times 10^{-8}$. The statistical uncertainty is only 15\% larger than expected from the neutron beam shot noise~\cite{Fry}.

\paragraph{Systematic uncertainties.}

Table 1 lists the largest systematic uncertainties in our measurement of $\Ag$. The variation in thickness of the formed aluminum entrance windows leads to an uncertainty in the fractional yield of prompt aluminum $\gamma$s, resulting in a systematic uncertainty in $\Ag$ of $1 \times 10^{-9}$ ~\cite{BlythAlPRC}. The targets used to measure the aluminum asymmetry were centered in the detector array, while the aluminum components of the apparatus were located near the upstream end of the detector. We tested our ability to calculate geometric factors for such different geometries by measuring the large Cl asymmetry with targets in the center, front, and back of the detector~\cite{Fomin2018}. The spread in the extracted Cl asymmetries was 3\%, which yields an additional uncertainty from the contribution of prompt aluminum $\gamma$'s of $7 \times 10^{-10}$. 

Another systematic uncertainty arises from bremsstrahlung $\gamma$'s from the $\beta$ decay of polarized $^{28}$Al. The $^{28}$Al ground state $\beta$ decays to the first excited state of $^{28}$Si and the direction of the $\beta$ and subsequent bremsstrahlung $\gamma$'s are correlated with the polarization direction by the PV $\beta$ asymmetry parameter, which is assumed to have its maximum possible value of unity. The bremsstrahlung yield was calculated from recent measurements~\cite{Pandola2015}. The spin-lattice relaxation of the polarized aluminum nuclei at room and LH$_2$ temperatures and the effects of the different polarization reversal patterns were included. The estimated systematic uncertainty was below $0.9 \times 10^{-10}$. 

All other systematic effects discussed in Ref.~\cite{gericke} were reconsidered and their limits were either unchanged or slightly reduced. False electronic asymmetries were periodically measured with the neutron beam off and light emitting diodes (LEDs) illuminating the scintillator crystals (LED ON) or not (LED OFF). False asymmetries in both cases were less than 1 $\times 10^9$.

\begin{table}{}
\begin{center}
\caption{Dominant sources of systematic uncertainty and their contributions to $\Ag$.}
\label{tab.systematic}
\begin{tabular}{| l | c |}
\hline
Source & Contribution  \\
\hline
\hline
Prompt Al $\gamma$s: window thickness & 1$\times10^{-9}$ \\
Prompt Al $\gamma$s: geometric factors & $7$$\times10^{-10}$ \\
$^{28}$Al bremsstrahlung & $<9$$\times10^{-11}$ \\
False electronic asymmetry (LEDs off) & $<1$$\times10^{-9}$ \\
False electronic asymmetry (LEDs on) & $<1$$\times10^{-9}$ \\
Remaining systematic uncertainty~\cite{gericke}& $<3\times10^{-10}$\\
\hline
\noindent Total & $<2 \times10^{-9}$\\
\hline
\end{tabular}
\end{center}
\end{table}

Multiplicative corrections are applied to the data to account for geometric factors and neutron polarization. These include the uncertainties in the neutron depolarization by orthohydrogen (1.6\%), geometric factors (3\%), beam polarization (0.5\%), and spin flipper efficiency (0.5\%). The relative uncertainties of the three analysis methods were estimated to be 1\%~\cite{Fry}. The combined uncertainty from these corrections is 3.6\%, which is negligible when added in quadrature with the 47\% statistical uncertainty in the PV asymmetry. 

The final result for the hydrogen asymmetry is $\Ag=[-3.0\pm 1.4(\text{stat})\pm 0.2(\text{sys})] \times 10^{-8}$. This is consistent with the statistics-limited phase 1 result and surpasses the precision of Ref.~\cite{Caviagnac1977} which was unable to resolve $\Ag$.


\begin{figure}[htpt] 
\includegraphics[width=0.45\textwidth]{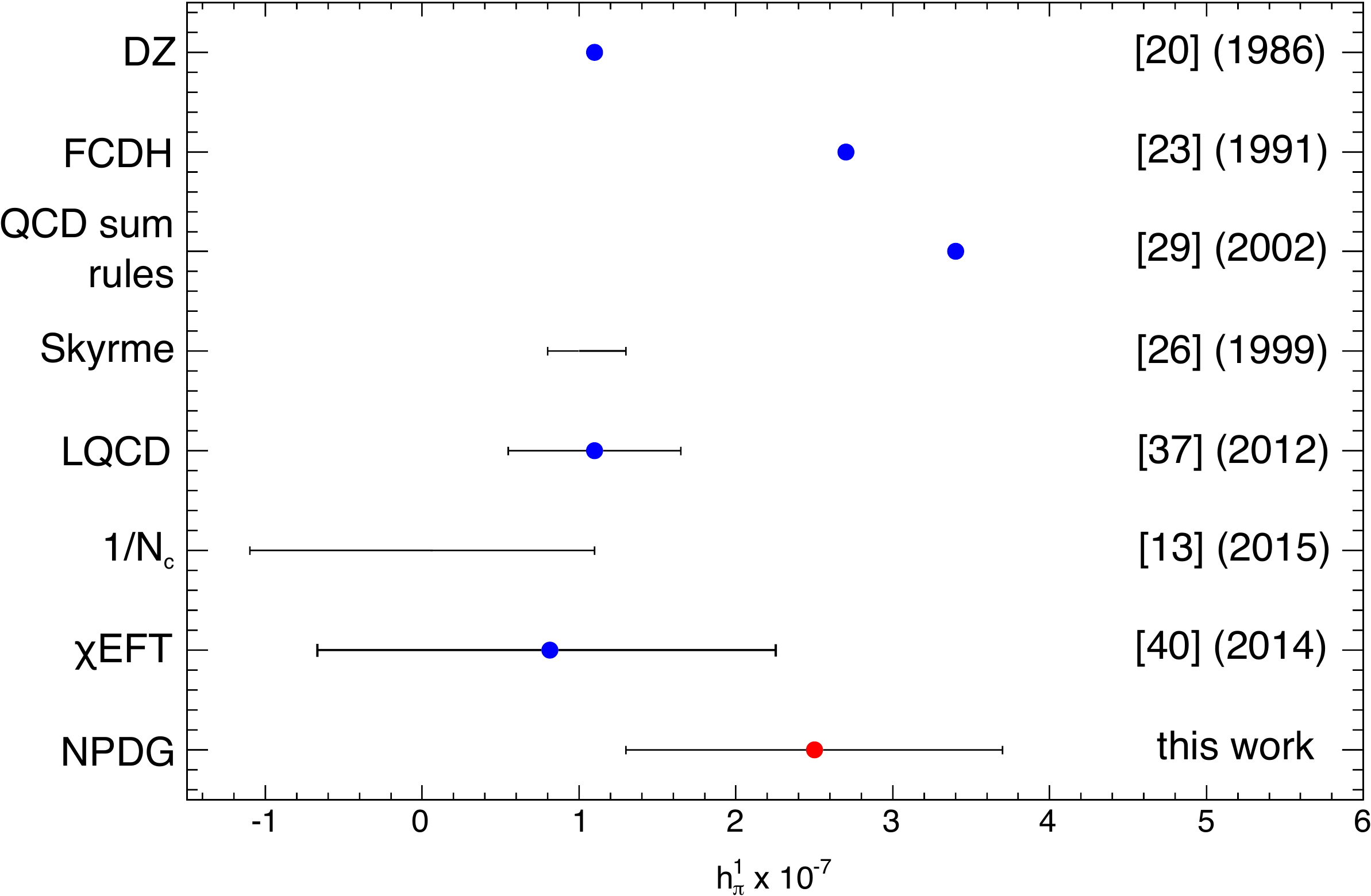}
\caption{$h_{\pi}^1$ from theoretical estimates or calculations (blue) and this work (red).}
\label{money}
\end{figure}

\paragraph{Discussion and Conclusion.}

We can extract a value of $h_{\pi}^{1}$  from the measured asymmetry because the heavy meson couplings enter the expression of $\Ag$ with very small coefficients. Hyun {\it et al.} ~\cite{Hyun2007} and Liu~\cite{Liu2007} give expansions of $\Ag$ in the meson-exchange picture using the AV18 NN potential: $\Ag = -0.117 h_{\pi}^{1} - 0.001 h_{\rho}^{1} + 0.002 h_{\omega}^{1}$ and $\Ag = -0.111 h_{\pi}^{1} - 0.001 h_{\rho}^{1} + 0.002 h_{\omega}^{1}$, respectively. We adopt the average of these two expansions, $\Ag = -0.114 h_{\pi}^{1} - 0.001 h_{\rho}^{1} + 0.002 h_{\omega}^{1}$. The RMS theoretical uncertainty in this procedure is 3\%, which is negligible compared to the statistical uncertainty. Neglecting heavy-meson terms, which contribute less than $1\%$ of $\Ag$ in the DDH reasonable range~\cite{Desplanques1980}, we obtain $h_{\pi}^{1} = [2.6 \pm 1.2(\text{stat}) \pm 0.2(\text{sys})] \times 10^{-7}$. Our value for $\Ag$ gives the pionless EFT coupling constant $C^{^{3}S_{1}\rightarrow ^{3}P_{1}}/C_{0}=[-7.4 \pm 3.5 (\text{stat}) \pm 0.5 (\text{sys})] \times 10^{-11}$ MeV$^{-1}$~\cite{Schindler2013}. Since $\Ag$ only depends on $h_{\pi}^1$ and $^{18}$F $P_{\gamma}$ contains all of the $\DI=1$ contributions, we can eliminate $h_{\pi}^1$ and find a constraint on the heavy mesons to be 0.4 $h_{\rho}^1$ + 0.6 $h_{\omega}^1$ = 8.5 $\pm$ 5.0, which is consistent with recent theoretical estimates ~\cite{Phillips2015,Gardner17}.

Figure \ref{money} shows an overview of theoretical estimates and this works extraction of  $h_{\pi}^1$. We report the most precise and direct determination of $h_{\pi}^1$ in a few-body system without atomic or nuclear corrections, and it is the best constraint for future investigation of the HWI. Additional theoretical and experimental work in exactly calculable few-body systems is needed to establish a complete determination of the HWI.

We gratefully acknowledge the support of the U.S. Department of Energy Office of Nuclear Physics through Grants DE-AC52-06NA25396, DE-FG02-03ER41258, DE-SC0014622, DE-AC-02-06CH11357, DE-AC05-00OR22725, and DE-SC0008107, the US National Science Foundation through Grants PHY-1306547, PHY-1306942, PHY-1614545, PHY-0855584, PHY-0855610, PHY-1205833, PHY-1506021, PHY-1205393, and PHY-0855694, PAPIIT-UNAM Grants IN110410, IN11193, and IG1011016, CONACYT Grant 080444, the Natural Sciences and Engineering Research Council of Canada (NSERC), and the Canadian Foundation for Innovation (CFI). This research used resources of the Spallation Neutron Source of Oak Ridge National Laboratory, a DOE Office of Science User Facility. J. Fry, R. C. Gillis, J. Mei, W. M. Snow, and Z. Tang acknowledge support from the Indiana University Center for Spacetime Symmetries. S.~Schr\"{o}der acknowledges support from the German Academic Exchange Service (DAAD).

\newpage{\pagestyle{empty}\cleardoublepage}

\end{document}